\documentstyle[12pt,titlepage]{article} 
\title{Estimates of B-Decays into K-Resonances and Dileptons}
\author{Mohammad R. Ahmady, Amir Fariborz and Dongsheng Liu}
\date{February, 1993}   
\def\_{\rule{.3em}{.15ex}}  

\begin{document}
\begin{titlepage}
 \begin{center}
  \vspace{0.75in}
  {\bf \LARGE Estimates of B-Decays into K-Resonances and Dileptons} \\
  \vspace{0.75in}
{\bf  Mohammad R. Ahmady}\\
Department of Physics, Ochanomizu University \\
 1-1, Otsuka 2, Bunkyo-ku, Tokyo 112, Japan\\
\vskip 1.0cm
{\bf Dongsheng Liu} \\
Department of Physics, University of Tasmania, Hobart, Australia 7001
\vskip 1.0cm
{\bf Amir H. Fariborz}\\
Department of Applied Mathematics, The University of Western Ontario \\
 London, Ontario, Canada N6A 5B7 \\

\vspace{1in}
  ABSTRACT \\
  \vspace{0.5in}
  \end{center}
  \begin{quotation}
  \noindent Short and long distance contributions to the
exclusive B-decays into various K-resonances
and dileptons, i.e. $B \rightarrow K^i \ell \bar \ell (\ell = e, \mu ,
\nu ) $, are examined.  The heavy quark effective theory has been used to
calculate the hadronic matrix elements.  Substantial branching fractions
are obtained for the dileptonic B-decays into some higher excited states
of K-mesons.  The long distance (resonance) contributions to these
exclusive rare B-decay modes dominate the short distance contributions
mostly by two orders of magnitude. It is pointed out that, excluding the
resonance contributions, the P-wave channels are dominant, accounting for
about $50\%$ of the inclusive $B\to X_s\ell^+\ell^-$ branching fraction.

  \end{quotation}
\end{titlepage}


\newcommand{\da}{\mbox{$\scriptscriptstyle \dag$}}
\newcommand{\lag}{\mbox{$\cal L$}}
\newcommand{\tr}{\mbox{\rm Tr\space}}
\newcommand{\fc}{\mbox{${\widetilde F}_\pi ^2$}}
\newcommand{\ns}{\textstyle}
\newcommand{\si}{\scriptstyle}

\section{Introduction}

Rare B-decays have been in the focus of a lot of theoretical attentions
.  This is due to the extensive amount
of information on Standard Model (SM) that can be extracted from these
processes.  The rare decays proceed through flavor
changing neutral current (FCNC) vertices that are absent at the tree level,
therefore, they are a good probe of the SM at the quantum (loop) level.  On
the other hand, radiative B-decays are sensitive to quark mixing
 angles
$V_{td}, V_{ts}$ and $ V_{tb}$,
hence their measurements yield valuable information
on CKM matrix elements, and consequently, shed some light on the CP
violation
in SM.  Also, reducing the uncertainties of the theoretical calculation
of these processes within the SM, and comparing them with more precise
measurements available at future B-factories, may be the most promising
probe of the new physics in the short term.

In this paper, we focus on the rare B-decays, $B \rightarrow K^i \ell \bar
 \ell (\ell = e, \mu , \nu ) $, where $K^i$ are various resonances
of K-meson.  We calculate both, the short-distance (SD), and the
long-distance (LD) contributions to these processes.
The motivation for this work is the following:  Due to the large phase
space available to rare B-decays, exclusive processes with excited
K-mesons in the final state might have substantial branching fractions.
In fact, in reference
\cite{AOM}, it is shown that for the rare B-decays $B \rightarrow
K^i \gamma$, which proceed through SD penguin operator, some higher
resonances of K-meson have 2-3 times larger branching ratios
than the ground state.  On the other hand, in reference \cite{ALT},
we indicated that the LD contributions to these processes are
very significant,  approximately 20\% of the SD amplitude.
Consequently, we feel that the same investigation should
 be carried out on dileptonic rare B-decays.

A main source of uncertainty, in all these estimates, has always been
the evaluation of hadronic matrix elements (HME) for specific exclusive
decays.  This would involve the long range nonperturbative QCD
effects which renders the calculations model-dependence.
We use the heavy quark effective theory (HQET)\cite{IW}, as a convenient
ansatz in formulating the HME.  This results in an enormous
simplification, especially when decay rates to higher excited K-meson
states are estimated.  With a plausible assumption that {\bf b} quark is
static within the B meson, one can show that these HME are the same as
those appearing in $B\to K^i\psi (\psi')$ decays.  Therefore, one can
adjust the universal form factor, such that a HQET formulation of the HME
fits the experimental data available for $B\to K^{(*)}\psi^{(\prime)}$\cite
{AL1}.  In fact, our predicted value for the ratio $R=\Gamma (B\to
K^*\gamma)/\Gamma (B\to X_s\gamma )$, obtained this way (see eqn. (22) in
Ref. \cite {AL1}), is in good agreement with the recent CLEO results \cite
{CLEO}.
In a follow up paper, we showed, by adopting a more general model for
the universal form factors, one can also estimate the non-leptonic decays
to higher K-meson resonances \cite {AL2}.  These predictions are yet to
be tested experimentally.

\section{Differential decay rates}
We start with the effective Lagrangian for $b\to s\ell^+\ell^-$, which
includes both, short and long distance contributions \cite {GSW,AMM}
\begin{equation}
\displaystyle
L_{eff} =\frac {G_F}{\sqrt 2} \left ( \frac {\alpha }{4 \pi s_W^2}
\right ) V^*_{ts}V_{tb}
(A\bar s L_\mu b \bar \ell L^\mu \ell +B \bar s L_\mu b \bar\ell
R^\mu \ell +2m_b s_W^2F \bar s T_\mu b \bar\ell \gamma^\mu\ell ),
\end{equation}
where
$$
L_\mu =\gamma_\mu (1-\gamma_5), \quad
R_\mu =\gamma_\mu (1+\gamma_5),
$$
and
$$
T_\mu = -i\sigma_{\mu \nu} (1+\gamma_5) q^\nu /q^2 .
$$
$V_{ij}$ are the Cabibbo-Kobayashi-Maskawa matrix elements,
$s_W^2=sin^2\theta_W\approx 0.23$ ($\theta_W$ is the weak angle), $G_F$
is the Fermi constant and $q$ is the total momentum of the final
$\ell^+\ell^-$ pair.

This effective Lagrangian is obtained by integrating out the heavy
degrees of freedom, ie. the top quark, W and Z bosons, at the scale $\mu
=M_W$.  Using the renormalization group equations to scale down to a
subtraction point comparable to the light masses, ie. $\mu \approx m_b$,
ensures
that large logarithms, $ln(M_W/m_b)$, are contained in the coefficient
functions A, B and F only.  On the other hand, these functions, which
their explicit forms can be found elsewhere \cite {GSW,AMM}, depend on
the top quark mass.  For our numerical evaluations we use $m_t=180$
GeV, which
is the weighted average of the recent CDF and D0 results \cite {CDFD0}.

The terms representing the long distance contributions to the decay
rates, enter coefficients $A$ and $B$ from the one-loop (charm) matrix
element of
the four-quark operators.  Besides $c\bar c$ continuum (free quark)
contribution, these coefficients receive the pole contributions from the
$J/\psi$ and $\psi'$, which due to their narrow widths, amount to
Breit-Wigner terms for these resonances.

  The HME of (1) are calculated by using the trace formalism in the
context of HQET \cite {FALK}:
\begin{equation}
<K^i(v^\prime )|\bar s \Gamma b|B(v)>=Tr \left [ \bar R^i(v^\prime
)\Gamma R(v)M(v, v^\prime )\right ].
\end{equation}
$R^i(v^\prime )$ and $R(v)$ are the matrix representations of $K^i$
and B respectively, $v$ and $v^\prime$ are velocities of initial and
final state mesons.  $M$, which represents the light degrees of
freedom, is related to the Isgur-Wise functions.

We classify various K-resonances into spin doublets \cite {AOM,AL2},
and thus, using (2), the HME to each doublet is expressed in terms
of a single universal function.  These functions represent the
underlying non-perturbative QCD dynamics, which at present is not
possible to calculate from first principles and certain model assumptions are
required for their evaluation.

We obtain the following expression for the differential decay
rates in the limit of massless leptons:
\begin{equation}
\begin{array}{rl}
\displaystyle\frac {d\Gamma (B\to K^i\ell^+\ell^- )}{dz} =&
\displaystyle
\frac{G_F^2}{96\pi^3}
\left ( \frac {\alpha}{4\pi s_W^2} \right )^2
M_B^5 y^2 {\left (x_+ x_-\right )}^{1/2}{\vert V_{ts}^*V_{tb}\vert}^2 \vert
\xi_I (x)\vert^2  \\ \times &
\displaystyle
\left  ( ( \vert  A \vert^2 +{\vert  B \vert}^2)F_i
          +2\vert  C \vert^2 G_i
          + 2Re[{( A + B)}^* C]H_i\right ), \\ &
\displaystyle
x=v\cdot v^\prime ,\; x_{\pm}=x\pm 1, \; y=\frac {m_{K^i}}{m_B} , \; z=\frac
{q^2}{m_B^2} ,\; C=s_W^2F ,
\end{array}
\end{equation}
where $\xi_I (x), (I=C, E, F$ and $G$), are the Isgur-Wise functions for
each spin doublet \cite {AOM,AL2}, and $q^2$ is the invariant mass of the
dilepton.  The functions $F_i$, $G_i$ and $H_i$ for various K-resonances are
tabulated in Table 1.  For $m_t=180$ GeV, $\alpha_s (M_W)=0.118$, and $\eta
=\alpha_s(\mu )/\alpha_s(M_W)$ for these processes taken to be $1.75$,
we obtain:
$$ \begin{array}{l}
A=2.020 +({\rm continuum+ resonance\; terms}),\\
B=-0.173 +({\rm continuum+ resonance \; terms}),\\
C=-0.146 .
\end{array}
$$
This shows, excluding the charm loop contributions, the coefficient $A$
is much larger than $B$ or $C$ for a large top quark mass.

Now we turn to the processes $B\to K^i \nu \bar \nu$, which
even though are  difficult to measure
 experimentally, can be tagged by
a large missing energy-momentum.  The effective Hamiltonian for $b\to
s\nu\bar\nu$, arises from the coupling of pure $V-A$ currents,
\begin{equation}
\displaystyle{ H_{eff} = - \frac {G_F}{\sqrt 2}
   \left (\frac {\alpha}{4\pi s_W^2}\right )
    V^*_{ts}V_{tb}[2E(t)] (\bar s L_\mu b) (\bar\nu L^\mu \nu),}
\end{equation}
where $t=m_t^2/m_W^2$.  The Wilson coefficient $E(t)$ (see ref. \cite {HWS}
for the explicit form), is not affected by QCD scaling corrections.  We
take $E(t)=1.66$ for $m_t=180$ GeV.  Using (2), we
obtain the following differential decay rates for $B\to K^i\nu\bar\nu$
\begin{equation}
\displaystyle {\frac {d\Gamma (B\to K^i\nu\bar\nu )}{dz} =\frac
{G_F^2}{{96\pi}^3}{\left (\frac {\alpha}{4\pi s_W^2}\right )}^2
M_B^5 y^2 {\left (x_+ x_-\right )}^{1/2}
{\vert V_{ts}^*V_{tb}\vert}^2 {\vert \xi_I (x)\vert }^2 }
\displaystyle {{\vert 2E(t)\vert}^2 (3F_i)}.
\end{equation}
The factor 3 mutiplying $F_i$ in (5), is due to the sum over
neutrino flavors.

\section{Isgur-Wise function and total decay rates}

In order to estimate the total
decay rates, $\Gamma (B\to
K^i\ell\bar\ell )$, we need to insert the Isgur-Wise functions, $\xi_I
(x)$, in (3) and (5). As we mentioned earlier, in doing so, we need to
adopt a model that could also be extended to higher excited final states.
The wavefunction model of
Isgur, Wise, Scora and Grinstein \cite {ISGW}, suits our purpose quite
well.  The functions
$\xi_I$, in this model, can be obtained from the overlap integrals:
\begin{equation}
\xi (v.v^\prime )= \sqrt {2L+1} i^L \int r^2 dr \Phi^* _F (r) \Phi_I (r)
j_L \left [ \Lambda r \sqrt {{(v.v^\prime )}^2 -1} \right ].
\end{equation}
I and F are initial and final radial wavefunctions
respectively, L is the orbital angular momentum of the final state meson,
and $j_L$ is the spherical Bessel function of order L.  The inertia parameter
$\Lambda$, is taken to be \cite{ALTO}
$$\Lambda = \frac{m_{K^i}m_q}{m_s+m_q},$$
where $m_q$ is the light quark mass.

  We follow IWSG model in using harmonic oscillator wavefunctions
   with oscillator strength $\beta$ for the radial wavefunctions $\Phi_I$ and
    $\Phi_F$.
$\beta$ is assumed to be the same for initial and final
     state mesons in order to satisfy the normalization condition $\xi_C (1
)=1$ and $\xi_{E,F,G} (1)=0$.   For example,
one obtains
\begin{equation}
\xi_C (v.v^\prime ) = exp \left [ \frac{9}{256 \beta^2} m_{K^i} ^2
(1- {(v. v^\prime )}^2) \right ]
\end{equation}
for the $(0^-,1^-)$ spin doublet by using the ground state radial
wavefunction. $\beta=0.295 GeV$ is fixed by the best fit of (7) to the
experimentally measured $B \rightarrow (K, K^* )+( \psi , \psi^\prime
)$ decays \cite{HFC,AL2}. We use the same value for $\beta$ in our
calculation of $\xi_E$, $\xi_F$, and$\xi_G$.

We would like to remark that eqn. (7) implies the dependence of the
universal form factor on the K-meson mass.  In the heavy quark limit, the
members of a spin doublet are degenerate in mass, which results in exactly
one form factor for each doublet.  However, for K-mesons, insertion of a
realistic value for the mass parameter in (7), leads to a different form
factor for each member of a doublet.  This can be thought of as a
relaxation of the spin symmetry to some extent which, in our view, results
in a more reliable estimate of $B\to K^i$ transitions.

The following values are used for our numerical estimates:
$$\begin{array}{ll}
m_B=5.28 {\rm GeV} & m_c=1.50 {\rm GeV}, \\
m_b=4.95 {\rm GeV} & \vert V_{ts}\vert =0.043. \\
\end{array}
$$
The partial decay widths to various K-resonances are tabulated in Table
2.  For $B\to K^i e^+e^-$ and $B\to K^i\mu^+\mu^-$, these partial decay
rates
are different only in the first interval, ie. small invariant dilepton mass
region, where the lower bound of the interval, $4m_\ell^2$, is determined by
the lepton mass.  This difference is shown in the first and second columns
of Table 2.  The enteries of the 3$^{\rm rd}$, 4$^{\rm th}$, 5$^{\rm th}$
and 6$^{\rm th}$ columns of that table are the partial decay widths for
$B\to K^i \ell^+\ell^-$ ($\ell =e\; ,\mu$) over
$\left ( {(m_\psi -\delta )}^2\; , {(m_\psi +\delta )}^2\right )$,
$\left ( {(m_\psi +\delta )}^2\; , {(m_{\psi'} -\delta )}^2\right )$,
$\left ( {(m_{\psi'} -\delta )}^2\; , {(m_{\psi'} +\delta )}^2\right )$,
and
$\left ( {(m_{\psi'} +\delta )}^2\; , {(m_B -m_{K^i} )}^2\right )$
 respectively.  We choose $\delta =0.2$ GeV for our numerical estimates.
For $K^*(1680)$ and $K_2(1580)$, $q^2_{\rm max}={(m_B-m_{K^i})}^2$ is in
fact smaller than ${(m_{\psi'}-\delta )}^2$ and ${(m_{\psi'}+\delta )}^2$
respectively.

For the decays to the ground state spin doublet ($K\; , \; K^*$), we have
also examined some other parametrizations of the Isgur-Wise function
which are commonly used in the literature.  These are the monopole form,
$$
\xi (v.v') = \frac {\omega_\circ^2}{\omega_\circ^2-2+2v.v'},
$$
with $\omega_\circ =1.8$, and the exponential form,
$$
\xi (v.v')=exp\left [\gamma (1-v.v')\right ],
$$
with $\gamma =0.5$.  The parameter values are obtained from the best
fit to the data on $D\to K\ell \nu_\ell$.

In Table 3, we compare the total branching fractions (excluding the
resonance contributions) when different Isgur-Wise functions are used for
dileptonic B-decays to $K$ and $K^*(892)$.  The total width of the B is
taken to be $\Gamma (B\to {\rm all})=5.6\times 10^{-13}$ GeV.

{}From (1) and (4), one can calculate the inclusive differential decay
rates, ie.

\begin{equation}
\begin{array}{rl}
\displaystyle\frac {1}{\Gamma (B\to X_c e\bar\nu )}\frac {d\Gamma}{dz} (B\to
X_s\ell^+\ell^- ) =& \displaystyle
\left ( \frac {\alpha}{4\pi s_W^2} \right )^2
\frac {2}{f(m_c/m_b)}
\frac {{\vert V_{ts}^*V_{tb}\vert}^2}{{\vert V_{cb}\vert}^2}
{(1-z)}^2 \\ \times &
\displaystyle
\left       ( ( \vert  A \vert^2 +{\vert  B \vert}^2)(1+2z)
                +2\vert  C \vert^2 (1+2/z)\right. \\
     \; &
\displaystyle
\left. +  6Re[{( A + B)}^* C]\right ) \\
\end{array}
\end{equation}
and
\begin{equation}
\begin{array}{rl}
\displaystyle\frac {1}{\Gamma (B\to X_c e\bar\nu )}\frac {d\Gamma}{dz} (B\to
X_s\nu\bar\nu ) =& \displaystyle
\left ( \frac {\alpha}{4\pi s_W^2} \right )^2
\frac {2}{f(m_c/m_b)}
\frac {{\vert V_{ts}^*V_{tb}\vert}^2}{{\vert V_{cb}\vert}^2}
{\vert 2E(t)\vert}^2\times 3 \\ \times &
\displaystyle
\left ( 1-3z^2+2z^3\right ),
\end{array}
\end{equation}
where
$$
f(x)=1-8x^2+8x^6-x^8-24x^4ln(x).
$$
Again, the factor 3 in (9) is due to sum over three neutrino
flavors.  By normalizing to the semileptonic rate in (8) and (9), the strong
dependence on the b-quark mass cancels out.  Using the above expressions,
and assuming $\vert V_{cb}\vert \approx\vert V_{ts}\vert$, we obtain:
\begin{equation}
\begin{array}{l}
\Gamma (B\to X_s e^+e^-)=4.15\times 10^{-18} {\rm GeV}, \\
\Gamma (B\to X_s\mu^+\mu^- )=2.81\times 10^{-18} {\rm GeV}, \\
\Gamma (B\to X_s\nu\bar\nu )=2.43\times 10^{-17} {\rm GeV},
\end{array}
\end{equation}
where the semileptonic branching ratio is taken to be $BR(B\to X_c
e\bar\nu )=0.105$.

In Table 4, we have tabulated the total decay rates (excluding the
resonance contributions) and the ratio $R=\Gamma(B\to
K^i\ell\bar\ell )/\Gamma(B\to X_s\ell\bar\ell )$ for various K-resonances.

\section{Concluding remarks}

We end this paper with a few concluding remarks.

{}From Table 2, we observe that the partial decay widths over the $\psi$
resonance region (the $3^{\rm rd}$ column) are about two orders of
magnitude larger than the short distance contributions which are mainly
from the first region (the $1^{\rm st}$ and $2^{\rm nd}$ columns for $e$
and $\mu$ respectively).  The exceptions are the d-wave transitions where
this dominance is reduced to roughly one order of magnitude.  As a
result, the total decay rates, $\Gamma (B\to K^i\ell^+\ell^-)$, are
dominated by the long distance (resonance) contributions.  The partial
decay widths over the $\psi'$ resonance region (the $5^{\rm th}$ column)
are not as large, and in fact, except for the ground state transitions,
the rest are mostly comparable or smaller than the short distance
contributions.

The most promising probe of the
short distance operators, which are sensitive to the new physics, can be
done by imposing an upper limit on the invariant mass of the
dileptons, ie. $q^2$ (see eqn. (3)), well below $m_\psi^2$.  As it is
reflected in the first and
second columns of Table 2, for some final state K-resonances, there is a
significant difference in this partial decay width between $e^+e^-$ and
$\mu^+\mu^-$ pair.  This difference is overshadowed in the total decay rate
 by the overwhelming long distance contribution which is independent of
the dilepton type.  On the other hand, much larger partial decay widths
 is observed for some higher excited K-meson states if the above upper
cut on $q^2$ is implemented.  For example, for $q^2\leq {(m_\psi
-\delta)}^2$,
$\Gamma (B\to K^*_2(1430)e^+e^-)$ and $\Gamma (B\to K^*_2(1430)\mu^+\mu^-)$
are larger than $\Gamma (B\to K^*e^+e^-)$ and $\Gamma (B\to K^*\mu^+\mu^-)$
 by factors of 6 and 4 respectively (see the $1^{\rm st}$ and $2^{\rm
nd}$ columns of Table 2).  As we mentioned in the introduction, this is
not unexpected, as there is a large mass difference between the initial
B-meson and the final state K-mesons.  Taking into account the efficiency
for the reconstruction of the decaying K-mesons \cite{ARGUS}, our
estimates in this work indicate that along with $K^*(892)$ and
$K_2^*(1430)$,
two other P-wave channels, ie. $K_1(1270)$ and $K_1(1400)$, are the best
possible modes to look for rare dileptonic B-decays in the future
B-factories.

We have compared our results for the ground state K-mesons, $K$ and
$K^*$, with those obtained from different parametrizations of the
Isgur-Wise function.  The branching fractions, tabulated in Table 3,
indicate that, in general, the monopole form of the Isgur-Wise function
results in a larger estimate of these decay modes.  We should also point
out that the smaller difference between $BR(B\to K^*e^+e^-)$ and $BR(B\to
K^*\mu^+\mu^-)$ in IWSG model, is the direct result of the large recoil
suppression of the Isgur-Wise function in this model.

Finally, we note from Table 4, that the exclusive channels we consider in
this paper, more or less saturate the dileptonic rare B-decay modes.
\\ \\
{\bf Acknowledgement}\\
The authors thank A. Ali, T. Zhijian and I. Watanabe for useful
discussions.  M.R.A. would like to acknowledge support from the Japanese
Society for the Promotion of Science.

\newpage
\vskip 2cm
\hskip -3.4cm
{\small
\begin{tabular}{c|c|c|c|c}
\hline
\multicolumn{1}{|c|}{$K^i$ Name} &
\multicolumn{1}{c|}{$ J^P $} &
\multicolumn{1}{c|}{$F^i$} &
\multicolumn{1}{c|}{$ G^i$} &
\multicolumn{1}{c|}{$H^i$}  \cr
\hline
\multicolumn{1}{|c|}{$K $} &
\multicolumn{1}{c|}{$0^-$} &
\multicolumn{1}{c|}{$x_+ x_-{(1+y)}^2$} &
\multicolumn{1}{c|}{$x_+ x_-$} &
\multicolumn{1}{c|}{$x_+ x_-(1+y)$} \cr
\hline
\multicolumn{1}{|c|}{$K^* (892)$} &
\multicolumn{1}{c|}{$1^-$} &
\multicolumn{1}{c|}{$x_+[x_+{(1-y)}^2+4xz]$} &
\multicolumn{1}{c|}{$x_+[x_++4/z({(1+y)}^2x_-+z)]$} &
\multicolumn{1}{c|}{$x_+[3(1-y)x_++2(1+y)x_-]$} \cr
\hline
\multicolumn{1}{|c|}{$K^* (1430)$} &
\multicolumn{1}{c|}{$0^+$} &
\multicolumn{1}{c|}{$x_+ x_-{(1-y)}^2$} &
\multicolumn{1}{c|}{$x_+ x_-$} &
\multicolumn{1}{c|}{$x_+ x_-(1-y)$} \cr
\hline
\multicolumn{1}{|c|}{$K _1 (1270)$} &
\multicolumn{1}{c|}{$1^+$} &
\multicolumn{1}{c|}{$x_-[x_-{(1+y)}^2+4xz]$} &
\multicolumn{1}{c|}{$x_-[x_-+4/z({(1-y)}^2x_+-z)]$} &
\multicolumn{1}{c|}{$x_-[3(1+y)x_-+2(1-y)x_+]$} \cr
\hline
\multicolumn{1}{|c|}{$K_1 (1400)$} &
\multicolumn{1}{c|}{$1^+$} &
\multicolumn{1}{c|}{$2/3x_-{x_+}^2[x_+{(1-y)}^2+xz]$} &
\multicolumn{1}{c|}{$2/3x_-{x_+}^2[x_-+1/z({(1-y)}^2x_++z)]$} &
\multicolumn{1}{c|}{$2/3x_-{x_+}^2[x_-(1+y)+x_-]$} \cr
\hline
\multicolumn{1}{|c|}{$K^* _2 (1430)$} &
\multicolumn{1}{c|}{$2^+$} &
\multicolumn{1}{c|}{$2/3x_-{x_+}^2[x_+{(1-y)}^2+3xz]$} &
\multicolumn{1}{c|}{$2/3x_-{x_+}^2[x_++3/z({(1-y)}^2x_+-z)]$} &
\multicolumn{1}{c|}{$2/3x_-{x_+}^2[x_+(1-y)+3x_-]$} \cr
\hline
\multicolumn{1}{|c|}{$K^* (1680)$} &
\multicolumn{1}{c|}{$1^-$} &
\multicolumn{1}{c|}{$2/3x_+{x_-}^2[x_-{(1+y)}^2+xz]$} &
\multicolumn{1}{c|}{$2/3x_+{x_-}^2[x_++1/z({(1+y)}^2x_--z)]$} &
\multicolumn{1}{c|}{$2/3x_+{x_-}^2[x_+(1-y)+x_-]$} \cr
\hline
\multicolumn{1}{|c|}{$K_2 (1580)$} &
\multicolumn{1}{c|}{$2^-$} &
\multicolumn{1}{c|}{$2/3x_+{x_-}^2[x_-{(1+y)}^2+3xz]$} &
\multicolumn{1}{c|}{$2/3x_+{x_-}^2[x_-+3/z({(1+y)}^2x_-+z)]$} &
\multicolumn{1}{c|}{$2/3x_+{x_-}^2[x_-(1+y)+3x_-]$} \cr
\hline
\multicolumn{1}{|c|}{$K(1460)$} &
\multicolumn{1}{c|}{$0^-$} &
\multicolumn{1}{c|}{$x_+ x_-{(1+y)}^2$} &
\multicolumn{1}{c|}{$x_+ x_-$} &
\multicolumn{1}{c|}{$x_+ x_-(1+y)$} \cr
\hline
\multicolumn{1}{|c|}{$K^* (1410)$} &
\multicolumn{1}{c|}{$1^-$} &
\multicolumn{1}{c|}{$x_+[x_+{(1-y)}^2+4xz]$} &
\multicolumn{1}{c|}{$x_+[x_++4/z(({(1+y)}^2x_-+z)]$} &
\multicolumn{1}{c|}{$x_+[3(1-y)x_++2(1+y)x_-]$} \cr
\hline
\end{tabular}

{\hskip -3.4cm Table 1: The functions $F_i$, $G_i$ and $H_i$ for various
K-resonances.}

\hskip 3.4cm

\vskip 2cm
\hskip -3.4cm
{\small
\begin{tabular}{c|c|c|c|c|c|c}
\hline
\multicolumn{1}{|c|}{$K^i$ Name} &
\multicolumn{1}{c|}{{\scriptsize $4m_e^2 , {(m_\psi -\delta )}^2 $}} &
\multicolumn{1}{c|}{{\scriptsize $4m_\mu^2 , {(m_\psi -\delta )}^2 $}} &
\multicolumn{1}{c|}{{\scriptsize ${(m_\psi -\delta)}^2 , {(m_\psi +\delta
)}^2 $}} &
\multicolumn{1}{c|}{{\scriptsize ${(m_\psi +\delta)}^2 ,  {(m_{\psi'}
-\delta )}^2 $}} &
\multicolumn{1}{c|}{{\scriptsize ${(m_{\psi'} -\delta)}^2 , {(m_{\psi'}
+\delta )}^2 $}} &
\multicolumn{1}{c|}{{\scriptsize ${(m_{\psi'} +\delta)}^2 ,  q_{\rm
max}^2$}} \cr \hline
\multicolumn{1}{|c|}{$K$} &
\multicolumn{1}{c|}{$1.1\times 10^{-19}$} &
\multicolumn{1}{c|}{$1.1\times 10^{-19}$} &
\multicolumn{1}{c|}{$3.3\times 10^{-17}$} &
\multicolumn{1}{c|}{$6.2\times 10^{-20}$} &
\multicolumn{1}{c|}{$2.4\times 10^{-18}$} &
\multicolumn{1}{c|}{$5.6\times 10^{-20}$} \cr
\hline
\multicolumn{1}{|c|}{$K^* (892)$} &
\multicolumn{1}{c|}{$2.1\times 10^{-19}$} &
\multicolumn{1}{c|}{$1.9\times 10^{-19}$} &
\multicolumn{1}{c|}{$8.0\times 10^{-17}$} &
\multicolumn{1}{c|}{$1.8\times 10^{-19}$} &
\multicolumn{1}{c|}{$8.3\times 10^{-18}$} &
\multicolumn{1}{c|}{$1.8\times 10^{-19}$} \cr
\hline
\multicolumn{1}{|c|}{$K^* (1430)$} &
\multicolumn{1}{c|}{$8.1\times 10^{-20}$} &
\multicolumn{1}{c|}{$8.0\times 10^{-20}$} &
\multicolumn{1}{c|}{$9.1\times 10^{-18}$} &
\multicolumn{1}{c|}{$6.6\times 10^{-21}$} &
\multicolumn{1}{c|}{$2.7\times 10^{-20}$} &
\multicolumn{1}{c|}{$3.2\times 10^{-23}$} \cr
\hline
\multicolumn{1}{|c|}{$K _1 (1270)$} &
\multicolumn{1}{c|}{$2.9\times 10^{-19}$} &
\multicolumn{1}{c|}{$1.8\times 10^{-19}$} &
\multicolumn{1}{c|}{$2.7\times 10^{-17}$} &
\multicolumn{1}{c|}{$2.4\times 10^{-20}$} &
\multicolumn{1}{c|}{$2.6\times 10^{-19}$} &
\multicolumn{1}{c|}{$7.3\times 10^{-22}$} \cr
\hline
\multicolumn{1}{|c|}{$K_1 (1400)$} &
\multicolumn{1}{c|}{$7.2\times 10^{-19}$} &
\multicolumn{1}{c|}{$5.7\times 10^{-19}$} &
\multicolumn{1}{c|}{$5.0\times 10^{-17}$} &
\multicolumn{1}{c|}{$3.8\times 10^{-20}$} &
\multicolumn{1}{c|}{$1.6\times 10^{-19}$} &
\multicolumn{1}{c|}{$2.3\times 10^{-22}$} \cr
\hline
\multicolumn{1}{|c|}{$K^* _2 (1430)$} &
\multicolumn{1}{c|}{$1.2\times 10^{-18}$} &
\multicolumn{1}{c|}{$7.5\times 10^{-19}$} &
\multicolumn{1}{c|}{$7.2\times 10^{-17}$} &
\multicolumn{1}{c|}{$5.3\times 10^{-20}$} &
\multicolumn{1}{c|}{$2.0\times 10^{-19}$} &
\multicolumn{1}{c|}{$2.3\times 10^{-22}$} \cr
\hline
\multicolumn{1}{|c|}{$K^* (1680)$} &
\multicolumn{1}{c|}{$3.3\times 10^{-20}$} &
\multicolumn{1}{c|}{$2.1\times 10^{-20}$} &
\multicolumn{1}{c|}{$1.7\times 10^{-19}$} &
\multicolumn{1}{c|}{$5.8\times 10^{-23}$} &
\multicolumn{1}{c|}{        -           } &
\multicolumn{1}{c|}{        -           } \cr
\hline
\multicolumn{1}{|c|}{$K_2 (1580)$} &
\multicolumn{1}{c|}{$1.1\times 10^{-19}$} &
\multicolumn{1}{c|}{$5.7\times 10^{-20}$} &
\multicolumn{1}{c|}{$1.1\times 10^{-18}$} &
\multicolumn{1}{c|}{$4.4\times 10^{-22}$} &
\multicolumn{1}{c|}{$       0          $} &
\multicolumn{1}{c|}{        -           } \cr
\hline
\multicolumn{1}{|c|}{$K(1460)$} &
\multicolumn{1}{c|}{$6.7\times 10^{-20}$} &
\multicolumn{1}{c|}{$6.7\times 10^{-20}$} &
\multicolumn{1}{c|}{$5.0\times 10^{-18}$} &
\multicolumn{1}{c|}{$2.8\times 10^{-21}$} &
\multicolumn{1}{c|}{$2.2\times 10^{-21}$} &
\multicolumn{1}{c|}{$1.3\times 10^{-24}$} \cr
\hline
\multicolumn{1}{|c|}{$K^* (1410)$} &
\multicolumn{1}{c|}{$2.5\times 10^{-19}$} &
\multicolumn{1}{c|}{$1.8\times 10^{-19}$} &
\multicolumn{1}{c|}{$3.1\times 10^{-17}$} &
\multicolumn{1}{c|}{$2.5\times 10^{-20}$} &
\multicolumn{1}{c|}{$1.3\times 10^{-19}$} &
\multicolumn{1}{c|}{$1.7\times 10^{-22}$} \cr
\hline
\end{tabular}

{\hskip -3.4cm Table 2: Partial decay widths (in GeV) for $B\to
K^i\ell^+\ell^-$ ($\ell
=e ,\mu$).  The difference between electron and muon }

{\hskip -1.9cm final states is in
the first interval only (small invariant mass for dilepton pair) which
is shown in}

{\hskip -1.9cm the first two columns.}

\hskip 3.4cm

\newpage
\vskip 2cm
\hskip -2.5cm
{\small
\begin{tabular}{c|c|c|c|c|c|c}
\hline
\multicolumn{1}{|c|}{$\begin{array}{l} {\rm Isgur-Wise} \\ {\rm
function} \end{array}$} &
\multicolumn{1}{c|}{$B\to Ke^+e^- $} &
\multicolumn{1}{c|}{$B\to K\mu^+\mu^- $} &
\multicolumn{1}{c|}{$B\to K\nu\bar\nu$} &
\multicolumn{1}{c|}{$B\to K^* e^+e^-$} &
\multicolumn{1}{c|}{$B\to K^*\mu^+\mu^-$} &
\multicolumn{1}{c|}{$B\to K^*\nu\bar\nu$} \cr
\hline
\multicolumn{1}{|c|}{$\begin{array}{l} {\rm Harmonic} \\ {\rm
wavefunction} \\
\beta =0.295 \end{array}$} &
\multicolumn{1}{c|}{$3.4\times 10^{-7}$} &
\multicolumn{1}{c|}{$3.4\times 10^{-7}$} &
\multicolumn{1}{c|}{$2.9\times 10^{-6}$} &
\multicolumn{1}{c|}{$8.9\times 10^{-7}$} &
\multicolumn{1}{c|}{$8.6\times 10^{-7}$} &
\multicolumn{1}{c|}{$8.0\times 10^{-6}$} \cr
\hline
\multicolumn{1}{|c|}{$\begin{array}{l} {\rm Monopole} \\ \omega
=1.8\end{array}$} &
\multicolumn{1}{c|}{$6.6\times 10^{-7}$} &
\multicolumn{1}{c|}{$6.6\times 10^{-7}$} &
\multicolumn{1}{c|}{$5.7\times 10^{-6}$} &
\multicolumn{1}{c|}{$3.8\times 10^{-6}$} &
\multicolumn{1}{c|}{$2.5\times 10^{-6}$} &
\multicolumn{1}{c|}{$2.1\times 10^{-5}$} \cr
\hline
\multicolumn{1}{|c|}{$\begin{array}{l} {\rm Exponential} \\ \gamma =
0.5\end{array}$} &
\multicolumn{1}{c|}{$2.9\times 10^{-7}$} &
\multicolumn{1}{c|}{$2.9\times 10^{-7}$} &
\multicolumn{1}{c|}{$2.5\times 10^{-6}$} &
\multicolumn{1}{c|}{$3.0\times 10^{-6}$} &
\multicolumn{1}{c|}{$2.1\times 10^{-6}$} &
\multicolumn{1}{c|}{$2.0\times 10^{-5}$} \cr
\hline
\end{tabular}

\hskip 2.5cm

{\hskip -2.5cm Table 3: Comparison between the branching fractions of the
rare B-decays to the ground state K-mesons}

{\hskip -1.0cm (resonance contributions are excluded) for different
Isgur-Wise functions.}

\vskip 2cm
\hskip -2.5cm
{\small
\begin{tabular}{c|c|c|c|c|c}
\hline
\multicolumn{1}{|c|}{$K^i$ Name} &
\multicolumn{1}{c|}{$ J^P $} &
\multicolumn{1}{c|}{Mass (MeV)} &
\multicolumn{1}{c|}{$B\to K^i e^-e^+$} &
\multicolumn{1}{c|}{$ B \rightarrow K^i \mu^-\bar \mu^+ $} &
\multicolumn{1}{c|}{$B\to K^i \nu \bar\nu$} \cr
\hline
\multicolumn{1}{|c|}{$K $} &
\multicolumn{1}{c|}{$0^-$} &
\multicolumn{1}{c|}{$497.67 \pm 0.03$} &
\multicolumn{1}{c|}{$1.9\times 10^{-19}\; ,\; (4.6)$} &
\multicolumn{1}{c|}{$1.9\times 10^{-19}\; ,\; (6.8)$} &
\multicolumn{1}{c|}{$1.6\times 10^{-18}\; ,\; (6.6)$} \cr
\hline
\multicolumn{1}{|c|}{$K^* (892)$} &
\multicolumn{1}{c|}{$1^-$} &
\multicolumn{1}{c|}{$896.1 \pm 0.3$} &
\multicolumn{1}{c|}{$5.0\times 10^{-19}\; ,\; (12.0)$} &
\multicolumn{1}{c|}{$4.8\times 10^{-19}\; ,\; (17.1)$} &
\multicolumn{1}{c|}{$4.5\times 10^{-18}\; ,\; (18.5)$} \cr
\hline
\multicolumn{1}{|c|}{$K^* (1430)$} &
\multicolumn{1}{c|}{$0^+$} &
\multicolumn{1}{c|}{$1429 \pm 7$} &
\multicolumn{1}{c|}{$6.5\times 10^{-20}\; ,\; (1.4)$} &
\multicolumn{1}{c|}{$6.5\times 10^{-20}\; ,\; (2.3)$} &
\multicolumn{1}{c|}{$6.0\times 10^{-19}\; ,\; (2.5)$} \cr
\hline
\multicolumn{1}{|c|}{$K _1 (1270)$} &
\multicolumn{1}{c|}{$1^+$} &
\multicolumn{1}{c|}{$1270 \pm 10$} &
\multicolumn{1}{c|}{$2.7\times 10^{-19}\; ,\; (6.5)$} &
\multicolumn{1}{c|}{$1.6\times 10^{-19}\; ,\; (5.7)$} &
\multicolumn{1}{c|}{$1.3\times 10^{-18}\; ,\; (5.3)$} \cr
\hline
\multicolumn{1}{|c|}{$K_1 (1400)$} &
\multicolumn{1}{c|}{$1^+$} &
\multicolumn{1}{c|}{$1402 \pm 7$} &
\multicolumn{1}{c|}{$6.3\times 10^{-19}\; ,\; (15.2)$} &
\multicolumn{1}{c|}{$4.8\times 10^{-19}\; ,\; (17.1)$} &
\multicolumn{1}{c|}{$3.6\times 10^{-18}\; ,\; (14.8)$} \cr
\hline
\multicolumn{1}{|c|}{$K^* _2 (1430)$} &
\multicolumn{1}{c|}{$2^+$} &
\multicolumn{1}{c|}{$1425.4 \pm 1.3$} &
\multicolumn{1}{c|}{$1.1\times 10^{-18}\; ,\; (26.5)$} &
\multicolumn{1}{c|}{$6.3\times 10^{-19}\; ,\; (22.4)$} &
\multicolumn{1}{c|}{$4.5\times 10^{-18}\; ,\; (18.5)$} \cr
\hline
\multicolumn{1}{|c|}{$K^* (1680)$} &
\multicolumn{1}{c|}{$1^-$} &
\multicolumn{1}{c|}{$1714 \pm 20$} &
\multicolumn{1}{c|}{$3.1\times 10^{-20}\; ,\; (0.7)$} &
\multicolumn{1}{c|}{$1.9\times 10^{-20}\; ,\; (0.7)$} &
\multicolumn{1}{c|}{$1.5\times 10^{-19}\; ,\; (0.6)$} \cr
\hline
\multicolumn{1}{|c|}{$K_2 (1580)$} &
\multicolumn{1}{c|}{$2^-$} &
\multicolumn{1}{c|}{$\approx 1580$} &
\multicolumn{1}{c|}{$1.0\times 10^{-19}\; ,\; (2.4)$} &
\multicolumn{1}{c|}{$5.1\times 10^{-20}\; ,\; (1.8)$} &
\multicolumn{1}{c|}{$3.0\times 10^{-19}\; ,\; (1.2)$} \cr
\hline
\multicolumn{1}{|c|}{$K(1460)$} &
\multicolumn{1}{c|}{$0^-$} &
\multicolumn{1}{c|}{$\approx 1460$} &
\multicolumn{1}{c|}{$5.4\times 10^{-20}\; ,\; (1.3)$} &
\multicolumn{1}{c|}{$5.4\times 10^{-20}\; ,\; (1.9)$} &
\multicolumn{1}{c|}{$4.7\times 10^{-19}\; ,\; (1.9)$} \cr
\hline
\multicolumn{1}{|c|}{$K^* (1410)$} &
\multicolumn{1}{c|}{$1^-$} &
\multicolumn{1}{c|}{$1412 \pm 12$} &
\multicolumn{1}{c|}{$2.2\times 10^{-19}\; ,\; (5.3)$} &
\multicolumn{1}{c|}{$1.5\times 10^{-19}\; ,\; (5.3)$} &
\multicolumn{1}{c|}{$1.3\times 10^{-18}\; ,\; (5.3)$} \cr
\hline
\multicolumn{1}{|c}{} &
\multicolumn{1}{c}{} &
\multicolumn{1}{c|}{} &
\multicolumn{1}{c|}{Sum: $76.1\%$} &
\multicolumn{1}{c|}{Sum: $81.1\%$} &
\multicolumn{1}{c|}{Sum: $75.2\%$} \cr
\hline
\end{tabular}
\hskip 2.5cm

{\hskip -2.5cm Table 4: The total decay rates, in GeV (resonance
contributions excluded), and the ratio}

{\hskip -1.0cm $R=\Gamma (B\to K^i\ell\bar\ell
)/\Gamma (B\to X_s \ell\bar\ell )$ (in $\%$) for various K-resonances.}

\end{document}